\documentclass[onecolumn,a4paper,preprintnumbers,amsmath,amssymb,nofootinbib,floatfix, aps, 10pt]{revtex4}

\usepackage[dvips]{epsfig}
\usepackage[english]{babel}
\usepackage{bbm}
\usepackage{verbatim}
\usepackage[utf8]{inputenc}
\usepackage{array}

\usepackage{multirow}
\usepackage{amsfonts}

\usepackage{hyperref}
\usepackage[dvipsnames]{xcolor}
\hypersetup{colorlinks=true,linkbordercolor=Blue,linkcolor=Blue, citecolor=Blue}
\usepackage{subeqnarray}
\usepackage{indentfirst} 

\makeatletter
\renewcommand*\l@section{\@dottedtocline{1}{1.5em}{2.3em}}
\makeatother

\begin{document}

\noindent

\title{On the generalized spinor classification: Beyond the Lounesto's Classification}
\noindent
\author{C. H. Coronado Villalobos$^{1}$} \email{carlos.coronado@inpe.br}
\author{R. J. Bueno Rogerio$^{2}$} \email{rodolforogerio@unifei.edu.br}
\author{A. R. Aguirre$^{2}$} \email{alexis.roaaguirre@unifei.edu.br}
\author{D. Beghetto$^{3}$} \email{dino.beghetto@unesp.br}
\noindent

\affiliation{$^1$Instituto Nacional de Pesquisas Espaciais (INPE),\\
CEP 12227-010, S\~ao Jos\'e dos Campos, SP, Brazil.} 
\affiliation{$^{2}$Instituto de F\'isica e Qu\'imica, Universidade Federal de Itajub\'a - IFQ (UNIFEI), \\
Av. BPS 1303, CEP 37500-903, Itajub\'a - MG, Brazil.}
\affiliation{$^3$Universidade Estadual Paulista (UNESP) Faculdade de Engenharia, Guaratinguet\'a, Departamento de F\'isica e Qu\'imica\\
CEP 12516-410, Guaratinguet\'a, SP, Brazil}


\begin{abstract}
In this paper we advance into a generalized spinor classification, based on the so-called Lounesto's classification. The program developed here is based on an existing freedom on the spinorial dual structures definition, which, in a certain simple physical and mathematical limit, allows us to recover the usual Lounesto's classification. 
The protocol to be accomplished here gives full consideration in the understanding of the underlying mathematical structure, in order to satisfy the quadratic algebraic relations known as Fierz-Pauli-Kofink identities, and also to provide physical observables. As we will see, such identities impose a given restriction on the number of possible spinorial classes allowed in the classification.
We also expose a mathematical device known as \emph{Clifford's algebra deformation}, which ensures real spinorial densities and holds the Fierz-Pauli-Kofink quadratic relations.
\end{abstract}

\maketitle

\section{Introduction}
The well known Lounesto's spinor classification is a comprehensive and exhaustive categorization based on the bilinear covariants that discloses the possibility of a large variety of spinors, comprising regular and singular spinors which includes the cases of Dirac, Weyl, and Majorana as very particular spinors \cite{bonorapandora}.

Hundreds of textbooks usually show the dual structure for the fermionic spin one-half Dirac field (or also known as eigenspinors of parity operator), only by elucidating a given structure without at least showing explicitly how it actually emerges, or even mentioning the fact that it may not be unique. Now, if other physical fields describing relevant particles exist, one may naturally wonder if they must obey the same dual structure commonly used for Dirac spinors or not. In other words, is the Dirac structure so fundamental?. Both questions are rarely asked in the physics literature.
Undertaking a deep analysis of spinorial duals could help us to get closer to answer the above-mentioned questions. The algebraic theory of spinor duals makes use of the rich and well known structure of Clifford's algebras to specify all possible duals for arbitrary algebras of any dimension and space(time) signature \cite{juliodual}.  However, when the theory of the mass-dimension-one (Elko) spinors was proposed, it was necessary to revisit some fundamental aspects of the Quantum Field Theory, such as spinorial dual theory and (deformation of) the Clifford's algebras, always aiming to retrieve physical information \cite{aaca, bilineares}.

Elko spinor, proposed in \cite{jcap}, is a spin-${1/2}$ fermionic field endowed with mass dimension one, built upon a complete set of eigenspinors of the charge conjugation operator, which has the property of being neutral
with respect to gauge interactions. On their earlier formulation, these fields were quantum objects which carry a representation of subgroups of the Lorentz group $H\!O\!M\!(2)$ and $S\!I\!M\!(2)$ \cite{cohen}, and corresponding semi-direct extension encompassing translation. Recently, a redefinition in the spinor adjoint has lead to a theory endowed with full Lorentz (Poincar\`e) symmetry. The main features of this formulation, along with the theory of duals may be found in Ref.\cite{mdobook}.
As Elko spinor have mass dimension one, there is nothing that precludes the appearance of mass dimension one spinor  further in classes $(4)$ and $(6)$ in Lounesto's classification \cite{cavalcanticlassification}.

In the context of the bilinear structures, it is indeed important to pay attention to the subtleties of Clifford's algebra when associating real numbers to the bilinear covariants \cite{bilineares, lounestolivro}. All the protocol developed in this paper is based on two fundamental works written by Crawford \cite{crawford1,crawford2}, where he worked out several important formalizations concerning the bispinor algebra, developing a rigorous mathematical mechanism to obtain real bilinear covariants. For a more complete understanding of this subject the reader is also referred to \cite{generalfierz, nishifierz, takahashifierz}. 

In this communication we look for the underlying mathematical approach that allows us to define a spinorial dual structure, by making correct use of the Crawford mechanism to evaluate the bilinear covariants, using the correct composition law of the basis vector of the Clifford's algebra (also taking into account the Dirac normalization), and verifying if such forms satisfy the algebraic Fierz-Pauli-Kofink identities (henceforth, we will call FPK identities). In the case they do not satisfy FPK identities, we present a similar procedure based on the deformation of the Clifford's algebra, which allows us to ensure that the FPK identities hold. 

The paper is organized as follows: in the next section we define the basic concepts concerning the construction of the spinorial duals, and then we define a general dual structure. What comes next, for a book-keeping purpose, is a deep overview on the well-known Lounesto's classification, showing the main aspects of the classification and the underlying algebraic structure behind it. Thus, we present the proposed program and then, working with a general dual structure, we define a general spinor classification holding the same algebraic structure as Lounesto's classification does. Finally, we present a mathematical mechanism which ensures real quadratic forms, and also guarantees the FPK identities. In the last section, we conclude with some final remarks.

\section{Proem: A brief overview on Spinorial Duals}\label{secaoduais}
Spinors may be defined in several different ways. In the context of Clifford's algebra, the spinors are defined to be
elements of a left minimal ideal, whereas in the context of group theory we say that the spinors are carriers of the fundamental representation of the group. Spinors are used extensively in physics \cite{carmeli}, and it is widely accepted that they are more fundamental than tensors (when the spacetime itself is represented by a manifold endowed with a Riemannian - or Lorenzian - metric structure). 

The idea that the usual Dirac dual cannot be applied to every spinor is sharp enough to force the development of an accurate criteria in the formalization of spinor duals \cite{juliodual}. As asserted in \cite{alternative}, regarding to  the physical observables, authors in \cite{whereareelko} classify Elko spinors as type-5. However, the Elko norm is defined taking into account the Elko dual structure, and  then all the physical quantities should carry the same dual structure rather than the Dirac one. This fact suggests the necessity of constructing bilinear forms using the correct dual structure.  
Until the present moment, there are three well-established dual structures in the theory of spinors. They are: the Dirac dual, given by
\begin{equation}\label{1}
\bar{\psi}_{h}=[\Xi_{D}(\boldsymbol{p})\psi_{h}]^{\dag}\gamma_0, 
\end{equation}
where the lower index $h$ concerns the helicity, and the operator $\Xi_D(\boldsymbol{p})$ can be given as, $\Xi_{D}(\boldsymbol{p})=\mathbbm{1}$ or $\Xi_{D}(\boldsymbol{p})\neq\mathbbm{1}$ \cite{ryder, peskin, bjorken, nondirac}; 
and the Elko dual, given by \cite{juliodual,aaca}\footnote{The presence of the abstract $\mathcal{O}$ operator stands for the new operators $\mathcal{A}$ and $\mathcal{B}$, built in Ref \cite{aaca}. Such operators were recently proposed and they are responsible to ensure the Lorentz invariance and locality of the Elko.} 
\begin{eqnarray}\label{2}
\stackrel{\neg}{\lambda}^{S/A}_{h}=[\Xi_{E}(\boldsymbol{p})\lambda^{S/A}_{h}]^{\dag}\gamma_0\mathcal{O}, 
\end{eqnarray}
where the operator $\Xi_E(\boldsymbol{p})$ is responsible to change the spinor helicity of the Elko spinors \cite{mdobook}. In general, the $\Xi(\boldsymbol{p})$ operators, presented in (\ref{1}) and (\ref{2}), must satisfy certain conditions given in \cite{mdobook, vicinity}.

Furthermore, $\Xi(\boldsymbol{p})$ has to be idempotent, $\Xi^2(\boldsymbol{p})=\mathbbm{1}$, ensuring an invertible mapping \cite{vicinity}. Thus, we have the following possibilities: $h=h'$, for which $\Xi(\boldsymbol{p})=\mathbbm{1}$ and stands for the Dirac usual case, $\Xi(\boldsymbol{p})\neq\mathbbm{1}$ stands for the non-standard Dirac adjoint \cite{nondirac}, and finally $h\neq h'$ leading to a more involved operator present in the mass-dimension-one theory \cite{aaca}.
Thus, the purpose of the present paper is to invoke a mathematical procedure for determining the spinorial dual structure based on the general form $\stackrel{\sim}{\psi}=[\Xi_G(\boldsymbol{p})\psi]^{\dag}\gamma_0$, by analysing the bilinear forms and the related FPK identities.
First of all, if the dual structure provides bilinear forms which ensure the FKP quadratic relations, then it will be sufficient to have a well-defined theory. Otherwise, we will show a method of deformation of the Clifford's algebra so that FPK identities are respected by the bilinear forms obtained from the new dual structure.

\section{Basic conceptions on the Lounesto's classification and Spinorial densities}\label{secaolounesto}

Suppose $\psi$ to be a given spinor which belongs to a section of the vector bundle $\mathbf{P}_{Spin^{e}_{1,3}}(\mathcal{M})\times\, _{\rho}\mathbb{C}^4$ where $\rho$ stands for the entire representation space $D^{(1/2,0)}\oplus D^{(0,1/2)}$, or a given sector of such \cite{crawford1, crawford2}. The bilinear quantities associated to $\psi$ read
\begin{eqnarray}\label{covariantes}
\sigma =\bar{\psi}\psi,\; \omega = i\bar{\psi}\gamma_5\psi, \nonumber\\  
\mathbf{J} = \bar{\psi} \gamma_\mu \psi \theta^\mu, \;\; \mathbf{K}= -\bar{\psi} \gamma_{5} \gamma_\mu \psi \theta^\mu,  \\ 
 \mathbf{S}=  \bar{\psi} i \gamma_{\mu \nu} \psi \theta^\mu \wedge \theta^\nu, \nonumber 
\end{eqnarray} 
where $\gamma_5:=-i\gamma_{0123}$ and $\gamma_{\mu\nu} : = \gamma_{\mu}\gamma_{\nu}$. Denoting the Minkowski metric by $\eta_{\mu \nu}$, the set $\{\mathbbm{1},\gamma_{I}\}$ (where $I\in\{\mu, \mu\nu, \mu\nu\rho, {5}\}$ is a composed index) is a basis for the Minkowski spacetime ${\mathcal{M}}(4,\mathbb{C})$ satisfying  $\{\gamma_{\mu},\gamma _{\nu}\}=2\eta_{\mu \nu }\mathbbm{1}$, and $\bar{\psi}=\psi^{\dagger}\gamma_{0}$ stands for the adjoint spinor with respect to the Dirac dual. Here, we are considering the space-time metric given by diagonal($1, -1, -1, -1$). The elements $\{ \theta^\mu \}$ are the dual basis of a given inertial frame $\{ \textbf{e}_\mu \} = \left\{ \frac{\partial}{\partial x^\mu} \right\}$, with $\{x^\mu\}$ being the global spacetime coordinates. 
Through the analysis of the above forms it is possible to construct a specific classification, namely Lounesto's classification, which provides six disjoint classes for the fermionic spin one-half fields \cite{lounestolivro}. Commonly, it is always expected to relate (\ref{covariantes}) to physical observables. For a space-time of dimension $N = 2n$, the corresponding spinor, $\psi$, has $D=2^{n}$ complex components which necessarily satisfy a system of $(D-1)^2$ quadratic relations (FPK identities). From a mathematical point of view, these identities are a direct consequences of the completeness of the Dirac matrices. 

So, we shall emphasize that, exclusively in the Dirac theory, the above bilinear covariants have particular interpretations. In this context, the mass of the particle is related to $\sigma$, and the pseudo-scalar $\omega$ is relevant for parity-coupling (the pseudo-scalar quantity interacts with a pseudo-scalar meson $\pi^0$ preserving parity \cite{dasqft}). In addition, $\sigma$ appears as mass and self-interaction terms in the Lagrangian, whereas $\omega$, being $\mathcal{CP}$-odd, might probe $\mathcal{CP}$ features \cite{heisenberginterplay}. Beside that, the current four-vector $\mathbf{J}$ gives the current of probability, 
$\mathbf{K}$ is an axial vector current,
and $\mathbf{S}$ is associated with the distribution of intrinsic angular momentum. In fact, $eJ_0$ is the charge density, whereas $ec J_k (i, j, k = 1, 2, 3)$ are identified to the (electric) current density \cite{minogin}. The $\frac{\hbar}{2} K_\mu$ is interpreted as chiral current (spin density), conserved when $m=0$ \cite{bonorapandora}. And, the quantity $\frac{e\hbar}{2mc} S_{ij}$ is the magnetic moment density, while $\frac{e\hbar}{2mc} S_{0j}$ is the electric moment density \cite{roldaolivro, ryder, claude, ferrari2017}.

In the Dirac theory, the two currents $J_{\mu}$ and $K_{\mu}$ are the Noether currents corresponding to the two transformations \cite{peskin}
\begin{eqnarray}
\psi(x) &\rightarrow& \psi^{\prime}(x) = e^{i\alpha}\psi(x), 
\\ 
\psi(x)  &\rightarrow& \psi^{\prime}(x)= e^{i\alpha\gamma_5}\psi(x).
\end{eqnarray}
The first of these is a $U(1)$ symmetry of the Dirac lagrangian, and explicitly provides the Noether conserved current $J_{\mu}$.
The second, called the chiral transformation, is a symmetry of the derivative term in the lagrangian but not of the mass term. Thus, the Noether theorem confirms that the axial vector current is conserved only if $m=0$ \cite{peskin}. 

We stress that such labels are given accordingly to the way that each one of these quantities (the time-component and space component) behaves under Lorentz transformations (for more details see chapter 17 in reference \cite{interpretacaobilineares}). 
Besides,  all the above structures are only valid for spinors which have the same dual structure as Dirac spinor does.
These mathematical structures mandatorily obey the so-called FPK identities, given by \cite{baylis}
\begin{eqnarray}\label{fpkidentidades}
\boldsymbol{J}^2 = \sigma^2+\omega^2, \;\; J_{\mu}K^{\mu} = 0,\;\; \boldsymbol{J}^2 &= -\boldsymbol{K}^2 \nonumber\\
J_{\mu}K_{\nu}-K_{\mu}J_{\nu} = -\omega S_{\mu\nu} - \frac{\sigma}{2}\epsilon_{\mu\nu\alpha\beta}S^{\alpha\beta},
\end{eqnarray}
in which 
\begin{eqnarray}
\boldsymbol{J}^2 = (J_\mu \theta^\mu)(J^\nu  \textbf{e}_\nu) = J_\mu J^\mu,
\end{eqnarray}
where we have used the definition of the dual basis, $\theta^\mu(\textbf{e}_\nu)=\delta^\mu_\nu$, and similarly $\boldsymbol{K}^2 = K_\mu K^\mu$, both clearly being scalars. The above identities are fundamental not only for classification, but also to guarantee the inversion theorem\cite{crawford1}. Within the Lounesto's classification scheme, a non-vanishing $\mathbf{J}$ is crucial, since it enables to define the so called boomerang \cite{lounesto2001clifford}, which has an ample geometrical meaning in order to assert that there are precisely six different classes of spinors. This is a prominent consequence of the definition of a boomerang. As far as the boomerang is concerned, it is not possible to exhibit more than six types of spinors, according to the bilinear covariants. Indeed, Lounesto's spinor classification splits them into regular and singular spinors. The regular spinors are those having at least one of the bilinear covariants $\sigma$ and $\omega$ non-null. On the other hand, singular spinors have $\sigma = 0 =\omega$. Consequently the FPK identities are normally replaced by the more general conditions \cite{crawford1, crawford2}
\begin{eqnarray}\label{multi}
Z^{2}=4\sigma Z, \;\;\; Z\gamma_{\mu}Z=4J_{\mu}Z, \;\; Zi\gamma_{5}Z=4\omega Z, \nonumber\\ 
Z\gamma_{5}\gamma_{\mu}Z=4K_{\mu}Z,\;\;\; Zi\gamma_{\mu}\gamma_{\nu}Z=4S_{\mu\nu}Z,
\end{eqnarray}
If an arbitrary spinor $\xi$ satisfies $\xi^{*}\psi\neq 0$ and belongs to $\mathbb{C}\otimes \mathcal{C}\ell_{1,3}$, or equivalently if $\xi^{\dagger}\gamma_{0}\psi\neq 0 \in \mathcal{M}(4,\mathbb{C})$, it is possible to recover the original spinor $\psi$ from its aggregate $Z$. Such relation is given by $\psi = Z\xi$, and the aggregate reads
\begin{eqnarray}\label{multivetorz}
Z=\sigma+\mathbf{J}+i\mathbf{S}+\mathbf{K}\gamma_{5}-i\omega\gamma_{5}.
\end{eqnarray}
Hence, using (\ref{multivetorz}) and taking into account that we are dealing with singular spinors, it is straightforward to see that the aggregate can be recast as
\begin{eqnarray}
Z=\mathbf{J}(1+i\mathbf{s}+ih\gamma_{0123}), \label{ZB}
\end{eqnarray} 
where $\mathbf{s}$ is a space-like vector orthogonal to $\mathbf{J}$, and $h$ is a real number \cite{lounestolivro}. The multivector as expressed in (\ref{ZB}) is a boomerang. By inspecting the condition $Z^{2}=4\sigma Z$, we see that $Z^2=0$ for singular spinors. However, in order to the FPK identities to hold, it is also necessary that both conditions $\mathbf{J}^2=0$ and $(\mathbf{s}+h\gamma_{0123})^2=-1$ must be satisfied \cite{spinorrepresentation}.

As it can be seen, the physical requirement of reality can always be satisfied for the Dirac spinors bilinear covariants \cite{crawford1}, by a suitable deformation of the Clifford's basis leading to physical appealing quantities. However, the same assertion cannot be stated for the mass-dimension-one spinors, for which the FPK relations are violated \cite{bilineares}. We suspect that this fact is due to the new dual structure associated to these spinors. It is worth to mention that the main difference between the Crawford deformation \cite{crawford1, crawford2} and the one to be accomplished here is that in the former case, the spinors are understood as Dirac spinors, i.e., spinorial objects endowed with single helicity \cite{bilineares}, while the protocol that will be developed here is a general  procedure that can fit in any case.

\section{On the set-up of a generalized Spinor classification}\label{secaoespinores}\label{general}
A natural path to classify spinors resides on the Lounesto's classification. Such classification is built up taking into account the 16 bilinear forms, encompassing Dirac spinors, Type-4 spinors \cite{dualtipo4}, Majorana spinors 
(neutrino), and Weyl spinors 
(massless neutrino) \cite{lounestolivro}. This specific classification is based on geometric FPK identities, given in (\ref{fpkidentidades}) and (\ref{multi}), and displays exclusively six disjoint classes of spinors. This fact is due to the restriction imposed by the FPK identities. Then, it covers all the possibilities of spinors restricted by this geometrical constraint.     

The quantum field theory literature usually takes the Dirac dual as the standard one, with no suspicion or need of alternative dual structures as being potentially interesting \cite{juliodual}. However, the development of the theory of dark spinors presented the need for a review of dual structures \cite{aaca}.
A very important and peculiar feature that concerns to the Lounesto's classification lies in the fact that such classification takes into account (\emph{exclusively}) the Dirac dual structure. However, if spinors with a more involving dual structure exist, how they may be classified? Is it possible to fit them within Lounesto's classification? Maybe there is a naive freedom in Lounesto's classification which allows one to develop a more general classification, taking into account the appropriated dual structure. In this section we look towards to develop a generalized spinor classification which allows us, in a certain and appropriate mathematical limit, to recover the usual Lounesto's classification. For the sake of completeness, we will consider the same rigorous mathematical procedure and conditions as Lounesto did. However we will impose a general spinorial dual structure.   

Let $\psi$ be a given (algebraic) spinor which belongs to a section of the vector bundle $\mathbf{P}_{Spin^{e}_{1,3}}(\mathcal{M})\times\, _{\rho}\mathbb{C}^4$, where $\rho$ stands for the entire representation space $D^{(1/2,0)}\oplus D^{(0,1/2)}$. Such spinor can be described as follows
\begin{equation}
\psi = \left(\begin{array}{cccc}
a & b & c & d
\end{array} \right)^{T},
\end{equation}
where the components $a, b, c, d \in \mathbb{C}$. Unlike what was developed by Lounesto, here we extend the dual structure to cover all possibilities of spinorial duals. Now, let us define the new dual structure as %
\begin{equation}\label{newdualstructure}
\stackrel{\sim}{\psi} \stackrel{def}{=} [\Xi_{G}(\boldsymbol{p})\psi]^{\dag}\gamma_0,
\end{equation}
where the general $\Xi_{G}(\boldsymbol{p})$ operator is given by
\begin{eqnarray}
\Xi_{G}(\boldsymbol{p}) = \left( \begin{array}{ccc}
m_{11} & ... & m_{14} \\ 
\vdots & \ddots & \vdots \\ 
m_{41} & ... & m_{44}
\end{array}  \right),
\end{eqnarray} 
with the components $m_{ij}\in\mathbb{C}$. This operator, \emph{a priori}, must obey the following two main constraints: $\Xi^2_{G}(\boldsymbol{p})=\mathbbm{1}$, and also $\Xi^{-1}_{G}(\boldsymbol{p})$ indeed exists, ensuring an invertible map. Notice that if one imposes $\Xi_{G}(\boldsymbol{p})\equiv\mathbbm{1}$, then we recover the usual Lounesto's classification. Therefore, the general dual structure is defined as follows 
\begin{eqnarray}
\stackrel{\sim}{\psi} = \left(\begin{array}{cccc} 
\overline{am_{3,1}+bm_{{3,2}}+cm_{{3,3}}+dm_{{3,4}}}&\\
\overline{am_{{4,1}}+bm_{{4,2}}+cm_{{4,3}}+dm_{{4,4}}}&\\
\overline{am_{{1,1}}+bm_{{1,2}}+cm_{{1,3}}+dm_{{1,4}}}&\\
\overline{am_{{2,1}}+bm_{{2,2}}+cm_{{2,3}}+dm_{{2,4}}}
\end{array}
 \right)^{T},
\end{eqnarray}
where the overline stands for the complex conjugation. Now, by using this structure one can evaluate the 16 bilinear quantities
\begin{eqnarray}\label{newbilinear1}
\stackrel{\sim}{\sigma} = \stackrel{\sim}{\psi}\psi, \;\;  \stackrel{\sim}{\omega} = -i\stackrel{\sim}{\psi}\gamma_5\psi, \nonumber\\
 \boldsymbol{\stackrel{\sim}{J}}= \stackrel{\sim}{\psi}\gamma_{\mu}\psi\theta^{\mu}, \;\; \boldsymbol{\stackrel{\sim}{K}}= -\stackrel{\sim}{\psi}\gamma_5\gamma_{\mu}\psi\theta^{\mu}, \\ 
\boldsymbol{\stackrel{\sim}{S}} = i\stackrel{\sim}{\psi} \gamma_{\mu}\gamma_{\nu}\psi\theta^{\mu}\wedge\theta^{\nu}.\nonumber
\end{eqnarray}
We highlight that the operator $\Xi_{G}(\boldsymbol{p})$ is dimensionless. In such a way, it does not affect the units of the bilinear forms. So far, such spinorial densities allow us to define a generalized spinor classification, providing the first six disjoint classes with $\stackrel{\sim}{\textbf{J}}\neq 0$, and the remaining three with $\stackrel{\sim}{\textbf{J}}=0$ (as developed in \cite{jotanulo}). Then, it reads
\begin{enumerate}
  \item $\stackrel{\sim}{\sigma}\neq0$, $\quad\stackrel{\sim}{\omega}\neq0$, \hspace{0.5cm} $\stackrel{\sim}{\textbf{K}}\neq0,$ $\quad\stackrel{\sim}{\textbf{S}}\neq0$,
  \item $\stackrel{\sim}{\sigma}\neq0$, $\quad\stackrel{\sim}{\omega}=0$,\hspace{0.5cm} $\stackrel{\sim}{\textbf{K}}\neq0,$ $\quad\stackrel{\sim}{\textbf{S}}\neq0$,
  \item $\stackrel{\sim}{\sigma}=0$, $\quad\stackrel{\sim}{\omega}\neq0$,\hspace{0.5cm} $\stackrel{\sim}{\textbf{K}}\neq0,$ $\quad\stackrel{\sim}{\textbf{S}}\neq0$,
  \item $\stackrel{\sim}{\sigma}=0=\stackrel{\sim}{\omega},$ \hspace{0.5cm} $\stackrel{\sim}{\textbf{K}}\neq0,$ $\quad\stackrel{\sim}{\textbf{S}}\neq0$,
  \item $\stackrel{\sim}{\sigma}=0=\stackrel{\sim}{\omega},$ \hspace{0.5cm} $\stackrel{\sim}{\textbf{K}}=0,$ $\quad\stackrel{\sim}{\textbf{S}}\neq0$,
  \item $\stackrel{\sim}{\sigma}=0=\stackrel{\sim}{\omega},$ \hspace{0.5cm} $\stackrel{\sim}{\textbf{K}}\neq0,$ $\quad\stackrel{\sim}{\textbf{S}}=0$,
   \item $\stackrel{\sim}{\sigma}=0=\stackrel{\sim}{\omega},$ \hspace{0.5cm} $\stackrel{\sim}{\textbf{K}}\neq0,$ $\quad\stackrel{\sim}{\textbf{S}}\neq0$, \hspace{0.1cm} $\stackrel{\sim}{Z} = i(\stackrel{\sim}{\textbf{S}}+\stackrel{\sim}{\textbf{K}}\gamma_{0123})$,
  \item $\stackrel{\sim}{\sigma}=0=\stackrel{\sim}{\omega},$ \hspace{0.5cm} $\stackrel{\sim}{\textbf{K}}=0,$ $\quad\stackrel{\sim}{\textbf{S}}\neq0$,\hspace{0.5cm} $\stackrel{\sim}{Z} = i\stackrel{\sim}{\textbf{S}}$,
  \item $\stackrel{\sim}{\sigma}=0=\stackrel{\sim}{\omega},$ \hspace{0.5cm} $\stackrel{\sim}{\textbf{K}}\neq0,$ $\quad\stackrel{\sim}{\textbf{S}}=0$, \hspace{0.5cm} $\stackrel{\sim}{Z} = i\stackrel{\sim}{\textbf{K}}\gamma_{0123}$,
 \end{enumerate}
where we have three regular classes and six singular classes. 
For spinors respecting the Dirac dynamics,  $\stackrel{\sim}{\textbf{J}}$ will be the conserved current, whereas the last three classes describe spinors obeying only the Klein-Gordon equation \cite{jotanulo}. We stress that, up to date, classes with $\stackrel{\sim}{\textbf{J}}=0$ have only mathematical significance, since no associated physical entities have been observed yet.
Through an exhaustive analysis of the FPK geometric identities, we can also guarantee that these are the \emph{only} nine possible classes to be constructed.

Because of this, the Elko spinor \cite{bilineares} belongs to class-2 of the above classification, i.e., it  belongs to a regular class. It is important to highlight why we are looking for such brand new classification. 

 The need for a real interpretation of such bilinear densities arises at this particular point, and then the statement presented in section 11.1 of \cite{dasqft}, shall shed some light for the interpretation that we are searching for. 

Thus, for the general classification developed above, $\stackrel{\sim}{\sigma}$ still stands for the invariant length. Moreover, the four-vector $\stackrel{\sim}{\textbf{J}}$ represents the electric current density for charged particles, whereas for neutral particles it can be interpreted as the effective electromagnetic current four-vector\cite{giuntineutrino}. The bilinear $\stackrel{\sim}{\textbf{K}}$ shall be related with the spin alignment due to a coupling with matter or electromagnetic field. Finally, $\stackrel{\sim}{\textbf{S}}$ is related to the electromagnetic momentum density for charged particles. Although for neutral particles, we can infer that it may correspond to the momentum spin-density, or even it may represent spin precession (spin oscillation) in the presence of matter or electromagnetic fields \cite{studenikinneutrino2}. Such effect is caused by the neutral particle interaction with matter polarized by external magnetic field or, equivalently, by the interaction of the induced magnetic momentum of a neutral particle with the magnetic fields \cite[and references therein]{grigoneutrino, ahluwalianeutrino}. On the other hand, the physical interpretation of the bilinear $\stackrel{\sim}{\omega}$ is not clear enough for us.
The observations above are a hint towards the physical meaning of the bilinears in this general classification.

Finally, let us consider the behaviour of the bilinear forms for the spinor under charge conjugation. To accomplish such task, we use the protocol described in \cite{dasqft}, which is based on the antisymmetrization of the corresponding bilinears. In this case, the antisymmetrized bilinear forms can be defined by
\begin{equation}\label{bilinearanti}
(\bar{\psi}\Gamma_{i}\psi)^{anti}=\frac{1}{2}(\bar{\psi}\Gamma_{i}\psi - \psi^{T}\Gamma_{i}^{T}\bar{\psi}^{T}).
\end{equation}
As usual, for the Dirac spinors we have that $Cu_{\pm}=iv_{\pm}$ and $Cv_{\pm}=iu_{\pm}$, for Majorana $C\psi_M=\psi_M$ and for Elko $C\lambda^{S/A}=\pm\lambda^{S/A}$ \cite{mdobook}. Therefore, in the Dirac case, $J_{\mu}^{anti}$ matches $J_{\mu}$ given in (\ref{covariantes}), whereas for neutral particles like Majorana and Elko we have the following
\begin{equation}
J_{\mu}^{anti} = 0, \;\mbox{and}\; K_{\mu}^{anti} = 0.
\end{equation}
Then, Majorana and Elko fermions are neutral, and hence they cannot have any electromagnetic interaction, which is in agreement with what was discussed about the electric and magnetic charges, $e_{\mathbf{M}}$ and $q_{\mathbf{M}}$, in \cite{elkomonopole}.

\section{A Detour on the Clifford's algebra basis deformation: Dirac normalization and real spinorial densities issue}\label{secaodeformacao}
This section is reserved for a more careful and formal analysis regarding the general bilinear forms. Here we specify a protocol that should be followed if  the previous prescription somehow does not respect the FPK identities, or if the calculated bilinear forms are not real quantities. Then, we will deform the basis of Clifford's algebra by employing the Dirac normalization procedure \cite{crawford1}, in order to ensure real bilinear forms which satisfy properly the FPK identities.
Consider the well-known constitutive relation of the Clifford's algebra
\begin{equation}\label{eita}
\lbrace \gamma_{\mu}, \gamma_{\nu}\rbrace = 2\eta_{\mu\nu}\mathbbm{1}, \quad \mu,\nu = 0,1,2,...,N-1,
\end{equation}
where $\eta_{\mu\nu}$ is a $N=2n$ even-dimensional space-time metric, diagonal$(1,-1,...,-1)$. The generators of the Clifford's algebra are the identity $\mathbbm{1}$, the vectors $\gamma_\mu$ (commonly represented by square matrices), and products of the vector basis, given by \cite{crawford1},
\begin{equation}\label{gammatil}
\tilde{\gamma}_{\mu_{1}\mu_{2}...\mu_{N-M}} \equiv \frac{1}{M!}\epsilon_{\mu_{1}\mu_{2}...\mu_{N}}\gamma^{\mu_{N-M+1}}\gamma^{\mu_{N-M+2...}}\gamma^{\mu_{N}}.
\end{equation} 
As one can easily check, the lowest value of $M$ is two, which stands for the smallest possible combination, and the highest is $M=N$. Moreover, the elements that form the real Clifford's algebra basis are given by 
\begin{equation}\label{set}
\lbrace \Gamma_{i}\rbrace \equiv \lbrace \mathbbm{1}, \gamma_{\mu}, \tilde{\gamma}_{\mu_{1}\mu_{2}...\mu_{N-2}},..., \tilde{\gamma}_{\mu},\tilde{\gamma}\rbrace,  
\end{equation} 
where $\tilde{\gamma}\equiv \tilde{\gamma}_{\mu_{1}\mu_{2}...\mu_{N-N}}$.    

Now, the existence of the $\Xi_G(\boldsymbol{p})$-operator, appearing in the general definition of the spinorial dual structure, makes necessary to adequate the Clifford's algebra basis, in order to get the right appreciation of the FPK relations. We highlight that, for the Dirac spinor case, the set (\ref{set}) is suitable deformed in order to provide real bilinear covariants.

The first two main structures arising from the Clifford's algebra basis are defined as 
\begin{equation}\label{sig}
\stackrel{\sim}{\sigma} \equiv\stackrel{\sim}{\psi}\mathbbm{1}\psi, \;\;\mbox{and}\;\; \stackrel{\sim}{J}_{\mu} \equiv\stackrel{\sim}{\psi}\gamma_{\mu}\psi,
\end{equation}
where $\stackrel{\sim}{\psi}$ is defined in (\ref{newdualstructure}). The requirement of reality $\stackrel{\sim}{\sigma}=\stackrel{\sim}{\sigma}^\dag$ automatically leads to $\gamma_0 = \Xi_G(\boldsymbol{p})\gamma_0\Xi_G(\boldsymbol{p})$, since $\Xi_G^{2}(\boldsymbol{p})=\mathbbm{1}$. This constraint is satisfied, in such a way that (\ref{sig}) is real. Requiring the same condition on $\stackrel{\sim}{J}_{\mu}$ in (\ref{sig}), it leads to the following constraint $\gamma_0\gamma_{\mu} = \Xi_G^{\dag}(\boldsymbol{p})\gamma_{\mu}^{\dag}\gamma_0\Xi_G(\boldsymbol{p})$, which cannot be fulfilled by the Clifford's vector basis. We highlight that if one performs any change in $\gamma_{\mu}$, it may lead to a change in the constitutive relation of the Clifford's algebra (\ref{eita}), which in general leads to inconsistencies, and then we would be forced to give up on the present approach. Clearly, that modification must be discarded.  

Hereupon, we may deform the usual basis in order to redefine the bilinear covariants, in such a way that they satisfy the FPK relations. From equation (\ref{gammatil}), and by considering that the norm for the spinors must assume real values, we have
\begin{eqnarray*}  
\tilde{\gamma}_{\mu_{1}...\mu_{N-M}}= \frac{i^{M(M-1)/2}}{M!}\Xi_G(\boldsymbol{p})\epsilon_{\mu_{1}...\mu_{N}}\gamma^{\mu_{N-M+1}...}\gamma^{\mu_{N}}\Xi_G(\boldsymbol{p}).
\end{eqnarray*}
From the last equation, we can define the bispinor Clifford's algebra basis as in (\ref{set}). For instance, let us consider the four-dimensional space-time, i.e $N=4$. In this specific case the basis is given by 
\begin{eqnarray}
&&M = 4 \quad\Rightarrow\quad \tilde{\gamma} = -i \Xi_G(\boldsymbol{p})\gamma_5\Xi_G(\boldsymbol{p}), 
\\
&&M = 3 \quad\Rightarrow\quad \tilde{\gamma}_{\mu} = -\Xi_G(\boldsymbol{p})\gamma_5\gamma_{\mu}\Xi_G(\boldsymbol{p}),
\\
&&M = 2 \quad\Rightarrow\quad  \tilde{\gamma}_{\mu\nu} = i\Xi_G(\boldsymbol{p}) \gamma_{\mu}\gamma_{\nu} \Xi_G(\boldsymbol{p}).
\end{eqnarray}

After some algebra, it can be shown that the above modifications are sufficient
to assure that the FPK identities (\ref{fpkidentidades}) are fully satisfied. Notice that if one imposes $\Xi_G(\boldsymbol{p})=\mathbbm{1}$, we recover the usual Crawford deformation and the general bilinear forms  recast into (\ref{covariantes}). It is worth pointing out that the above deformation procedure does not necessarily provide a set of real bilinear quantities. All the forms in (\ref{covariantes}) were built upon spin one-half fermions under the Dirac dual structure. However, as it was shown, this may not be unique, and if more dual spinorial structures exist, then new classifications and bilinears forms must be constructed and analyzed.
  
\section{Final Remarks}\label{remarks}\label{secaoconclusao}
In the present communication we have developed a general spinor classification based on a general dual structure consistent with the FPK quadratic relations. The bilinear forms play a very important role, not only in classifying spinors but also to provide a physical interpretation. Thus, we have used the general procedure to interpret the physics related to the eigenspinors of parity operator and eigenspinors of charge conjugation operator. In fact, given a spinor related to some of the mentioned symmetries, such a relation will provide a physical interpretation of the bilinear forms, as the case of the Dirac, Majorana and Weyl spinors. Note that only a small group of spinors are physically relevant and then, the spinors which are not related to above mentioned symmetries should be analyzed in other scenarios, to ascertain whether they can possibly carry physical information, or only exist by algebraic construction.

As one can easily see, the Lounesto's classification is a very particular case of the one proposed in here. Such general classification encompasses exclusively nine classes. This restriction is due to the FPK algebraic relations, which can never be evaded. In this construction we look for accommodating all types of particles, both charged and neutral. Although we  have found a \emph{restricted} set of fermionic spin-$1/2$ particles that can be classified, we still seek to develop a classification where any other particle can be fitted. As highlighted in \cite{nogo}, a new range of possibilities still open windows to compose the standard model of high energy physics. This line of reasoning began when a new dual structure was presented in the literature, the Elko's dual structure. Thus, we wonder whether or not these spinors should be classified within Lounesto's classification, since they evade the dual spinorial structure imposed in such classification. Under these circumstances, we suppose that each spinor that has a different dual structure may present different interactions, different mass dimensionality, and then it needs an specific spinorial classification.

Regarding to the interpretation of the bilinear forms, as we have already mentioned in the course of the paper, for an exact understanding we need to know the mass dimension of the spinor, and evidently the possible couplings. In such a way, we would be able to infer about the physical interpretation of the bilinear by following the same path that Lounesto follows for the classification of Dirac spinors. As previously was thought, neutral particles do not carry any electric current or can not electromagnetically interact, so in this sense the current and the axial-current vanish in the antisymmetrization programme, as already expected.  Albeit still several studies related to neutral particles are in development and the results obtained are just simple observations, we expect that this may help us by bringing a coherent interpretation to the bilinear forms for such particles, and somehow making an association with the effective current density, momentum and spin precession in the presence of matter fields and even electromagnetic fields.
To conclude the program presented here, we also proposed a subtle deformation on the Clifford's algebra. Such a procedure is intended to use the general spinorial dual on the construction of the 16 basis elements. This mathematical procedure undergoes on a deformation of the basis of Clifford's algebra in order to provide new bilinear forms (via Dirac normalization method), leading to real spinorial densities, and also quantities that hold the FPK identities. Thus, such a mechanism is mathematically necessary if everything we have developed so far does not provide real spinorial densities, or does not satisfy FPK identities.

\section{Acknowledgements}
The authors express their gratitude to Rold\~ao da Rocha Jr, Julio Marny Hoff da Silva and Luca Fabbri for very stimulating and fruitful conversation, providing many insightful suggestions and questions. 
Authors also thanks to Marco Andr\'e Ferreira Dias and Jos\'e Abdalla Helay\"el-Neto for their generosity and for engaging in insightful discussions. CHCV thanks CNPq (PCI Grant Number 300381/2018-2) for the financial support. RJBR thanks to CNPq (Grant Number 155675/2018-4) for the financial support. ARA thanks Gabriel Flores Hidalgo for helpful discussions and comments on this essay.


\bibliographystyle{unsrt}
\bibliography{refs}

\end{document}